\documentclass[twocolumn]{aastex62}

\usepackage{natbib}
\usepackage{amsmath}
\usepackage{subfigure}
\setcitestyle{sort,comma}
\usepackage{csvsimple}

\usepackage{tabularx}
\definecolor{citecolor}{cmyk}{1,1,0,0}

\graphicspath{{./}{figures/}}

\shorttitle{Interior Analysis and Spitzer Secondary Eclipse Detection of TOI-824\,b}
\shortauthors{Roy et al.}

\begin{document}


\title{\large{Is the hot, dense sub-Neptune TOI-824\,b an exposed Neptune mantle?\\ \textit{Spitzer} detection of the hot day side and reanalysis of the interior composition}}



\correspondingauthor{Pierre-Alexis Roy}
\email{pierre-alexis.roy@umontreal.ca}

\author[0000-0001-6809-3520]{Pierre-Alexis Roy} 
\affil{Department of Physics and Institute for Research on Exoplanets, Universit\'{e} de Montr\'{e}al, Montreal, QC, Canada}

\author[0000-0001-5578-1498]{Bj\"{o}rn Benneke}
\affil{Department of Physics and Institute for Research on Exoplanets, Universit\'{e} de Montr\'{e}al, Montreal, QC, Canada}

\author[0000-0002-2875-917X]{Caroline Piaulet} 
\affil{Department of Physics and Institute for Research on Exoplanets, Universit\'{e} de Montr\'{e}al, Montreal, QC, Canada}

\author[0000-0002-1835-1891]{Ian J. M. Crossfield}
\affiliation{Department of Physics and Astronomy, University of Kansas, Lawrence, KS 66045}

\author[0000-0003-0514-1147]{Laura Kreidberg}
\affil{Max-Planck Institut f\"{u}r Astronomie, K\"{o}onigstuhl 17, 69117, Heidelberg, Germany}

\author[0000-0003-2313-467X]{Diana Dragomir}
\affil{Department of Physics and Astronomy, University of New Mexico, Albuquerque, NM, USA}

\author[0000-0001-5727-4094]{Drake Deming}
\affil{Department of Astronomy, University of Maryland, College Park, MD, USA}

\author[0000-0003-4990-189X]{Michael W. Werner}
\affil{Jet Propulsion Laboratory, California Institute of Technology, 4800 Oak Grove Drive, Pasadena, CA 91109, USA}

\author[0000-0001-9521-6258]{Vivien Parmentier}
\affil{Atmospheric, Oceanic \& Planetary Physics, Department of Physics, University of Oxford, Oxford OX1 3PU, UK}




\author[0000-0002-8035-4778]{Jessie L. Christiansen}
\affil{Caltech/IPAC-NExScI, M/S 100-22, 1200 E. California Blvd, Pasadena, CA 91125, USA}

\author[0000-0001-8189-0233]{Courtney D. Dressing}
\affiliation{Department of Astronomy, University of California, Berkeley, Berkeley, CA 94720}

\author[0000-0002-7084-0529]{Stephen R. Kane}
\affiliation{Department of Earth and Planetary Sciences, University of California, Riverside, CA 92521, USA}


\author[0000-0001-9414-3851]{Farisa Y. Morales}
\affiliation{Jet Propulsion Laboratory, California Institute of Technology, 4800 Oak Grove Drive, Pasadena, CA 91109, USA}

\begin{abstract}
The \textit{Kepler} and \textit{TESS} missions revealed a remarkable abundance of sub-Neptune exoplanets. Despite this abundance, our understanding of the nature and compositional diversity of sub-Neptunes remains limited, to a large part because atmospheric studies via transmission spectroscopy almost exclusively aimed for low-density sub-Neptunes and even those were often affected by high-altitude clouds. The recent \textit{TESS} discovery of the hot, dense TOI-824\,b ($2.93\,R_\oplus$ and $18.47\,M_\oplus$) opens a new window into sub-Neptune science by enabling the study of a dense sub-Neptune via secondary eclipses. Here, we present the detection of TOI-824\,b's hot day side via \textit{Spitzer} secondary eclipse observations in the $3.6$ and $4.5\,\mathrm{\mu m}$ channels, combined with a reanalysis of its interior composition. The measured eclipse depths (142$^{+57}_{-52}$ and 245$^{+75}_{-77}$\,ppm) and brightness temperatures (1463$^{+183}_{-196}$ and 1484$^{+180}_{-202}$\,K) indicate a poor heat redistribution ($f>$ 0.49) and a low Bond albedo (A$_{B}<$ 0.26). We conclude that TOI-824\,b could be an ``exposed Neptune mantle'': a planet with a Neptune-like water-rich interior that never accreted a hydrogen envelope or that subsequently lost it. The hot day-side temperature is then naturally explained by a high-metallicity envelope re-emitting the bulk of the incoming radiation from the day side. TOI-824\,b's density is also consistent with a massive rocky core that accreted up to 1\% of hydrogen, but the observed eclipse depths favor our high-metallicity GCM simulation to a solar-metallicity GCM simulation with a likelihood ratio of 7:1. The new insights into TOI-824\,b's nature suggest that the sub-Neptune population may be more diverse than previously thought, with some of the dense hot sub-Neptunes potentially not hosting a hydrogen-rich envelope as generally assumed for sub-Neptunes.

\end{abstract}


\section{Introduction} \label{sec:intro}

One of the great legacies of the \textit{Kepler} mission is the discovery that close-in small planets, especially sub-Neptunes, are abundant in the galaxy. For planetary systems with short orbital periods ($<$100 days), there are roughly 10 times more small exoplanets than large ones (larger than Neptune) \citep{fulton_california-_2018}. Furthermore, \textit{Kepler} has shown that the small planets are divided into two distinct populations: super-Earths and sub-Neptunes, separated by a lack of observed planets with size around $\sim$1.7-2\,$R_\mathrm{\oplus}$ and called the radius valley \citep{fulton_california-_2017, vaneylen_asteroseismic_2018, petigura_two_2020}. 

This bimodal distribution of super-Earths and sub-Neptunes has been the source of multiple studies of planet formation theory and many mechanisms have been proposed to explain the phenomenon. Photoevaporation, the process in which the X-ray and extreme UV (XUV) radiation from the host stars strips the H$_\mathrm{2}$/He envelope of young planets \citep{owen_kepler_2013} has been theorized to sculpt the population into these distinct categories : planets that still have a H$_\mathrm{2}$/He envelope (sub-Neptunes) and planets that lost it (super-Earths). Core-powered mass-loss, the process by which the internal energy from the planet's formation is enough to induce atmospheric escape is another mechanism that could explain the radius valley \citep{ginzburg_core-powered_2018}. Other models have proposed a ``primordial" bimodal distribution, created earlier during the formation of planetary cores and atmospheres, without needing active mass-loss or erosion later, during the planet's evolution \citep{lee_make_2014, lee_breeding_2016, lopez_how_2018, venturini_nature_2020, lee_primordial_2021}. While the origin of this radius valley is still under study, the \textit{Kepler} observations show that the sub-Neptunes, the planets on the larger end of the radius valley, outnumber the super-Earths, even when considering the overall greater detectability of sub-Neptunes \citep{fulton_california-_2018}.

Current consensus mostly describes sub-Neptunes as ``gas-rich super-Earths", i.e., planets that have rocky interiors (similar to super-Earths), but that accreted a large H$_2$-rich envelope and were able to keep it over their evolution \citep{bean_nature_2021}. The close-in formation of rocky cores inside of the ice line, or the migration of large planetary cores from beyond the ice line are two processes that are thought to lead to the formation of close-in super-Earths and sub-Neptunes (see Section \ref{sec:disc}). The final orbital position and stellar irradiation of the planet is then thought to discriminate between super-Earths (that lose their atmosphere to mass-loss processes) and sub-Neptunes (that keep their atmospheres), as was discussed above. Up until now, sub-Neptunes that have been characterized in transit or eclipse spectroscopy \citep[e.g., ][]{kreidberg_clouds_2014, benneke_water_2019, benneke_sub-neptune_2019} fit into this H$_2$-rich description of sub-Neptunes. We should however note that recent works have studied new mass-radius relations that can explain most sub-Neptunes with substantial water mass fractions \citep[e.g.,][]{mousis_irradiated_2020}.

Despite the large occurrence rate of sub-Neptunes, a detailed understanding of their formation and composition stays largely limited to date. First, the fact that there is no sub-Neptune in the solar system has prevented us from building a basic understanding of this type of planet. Second, virtually all sub-Neptunes with measured radii and masses occupy a degenerate region in mass-radius space, where their bulk composition cannot be determined from their size and mass alone \citep{rogers_three_2010, rogers_framework_2010}. Their properties are equally consistent with either rocky cores in hydrogen atmospheres or smaller cores in volatile-rich atmospheres with less hydrogen. Finally, atmospheric characterizations are challenging for sub-Neptunes, and even the low-density planets (with large atmosphere scale heights) favorable for transit observations are usually difficult to study because of clouds and aerosols. Most of the observations so far were obtained in transmission and the presence of clouds and aerosols in the atmospheres, especially in the case of lower-temperature planets, have muted parts of the spectral features \citep[e.g.,][]{knutson_hubble_2014, guo_updated_2020, libby-roberts_featureless_2020, bean_nature_2021}, a well-known example being the now infamous cloudy sub-Neptune GJ\,1214\,b \citep{bean_ground-based_2010, bean_optical_2011, desert_observational_2011, berta_flat_2012, fraine_spitzer_2013, kreidberg_clouds_2014}. We should note that studies of cooler targets, such as the habitable-zone sub-Neptune K2-18\,b, have been able to constrain the atmospheric abundances of molecules such as water via transmission spectroscopy \citep{benneke_water_2019}. However, outside of a few favorable targets, the detailed composition of sub-Neptunes remains mostly elusive to date.

\begin{figure*}[t]
  \centering
  \subfigure{\includegraphics[width=0.47\textwidth]{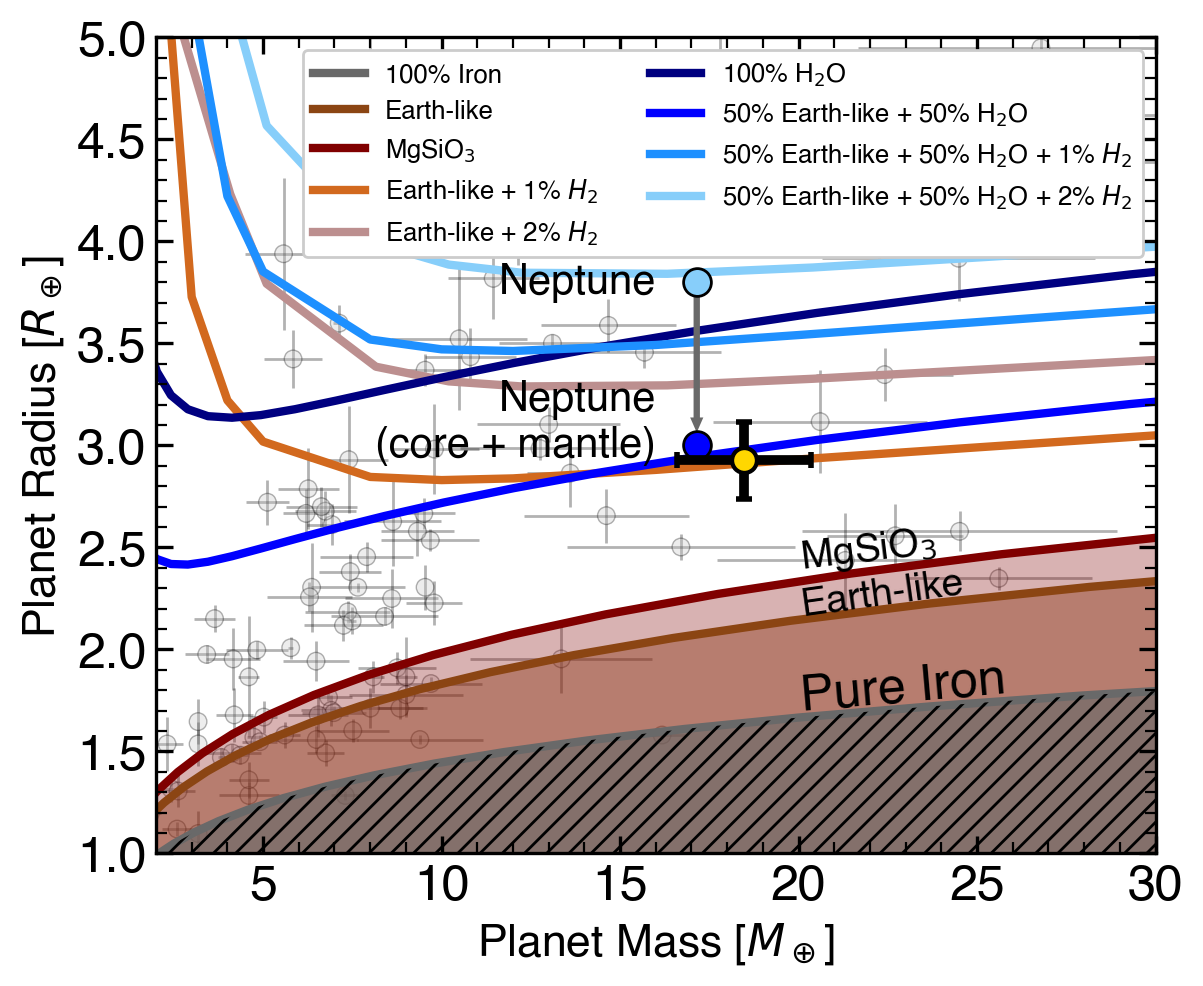}}
  \hfill
\subfigure{\includegraphics[width=0.485\textwidth]{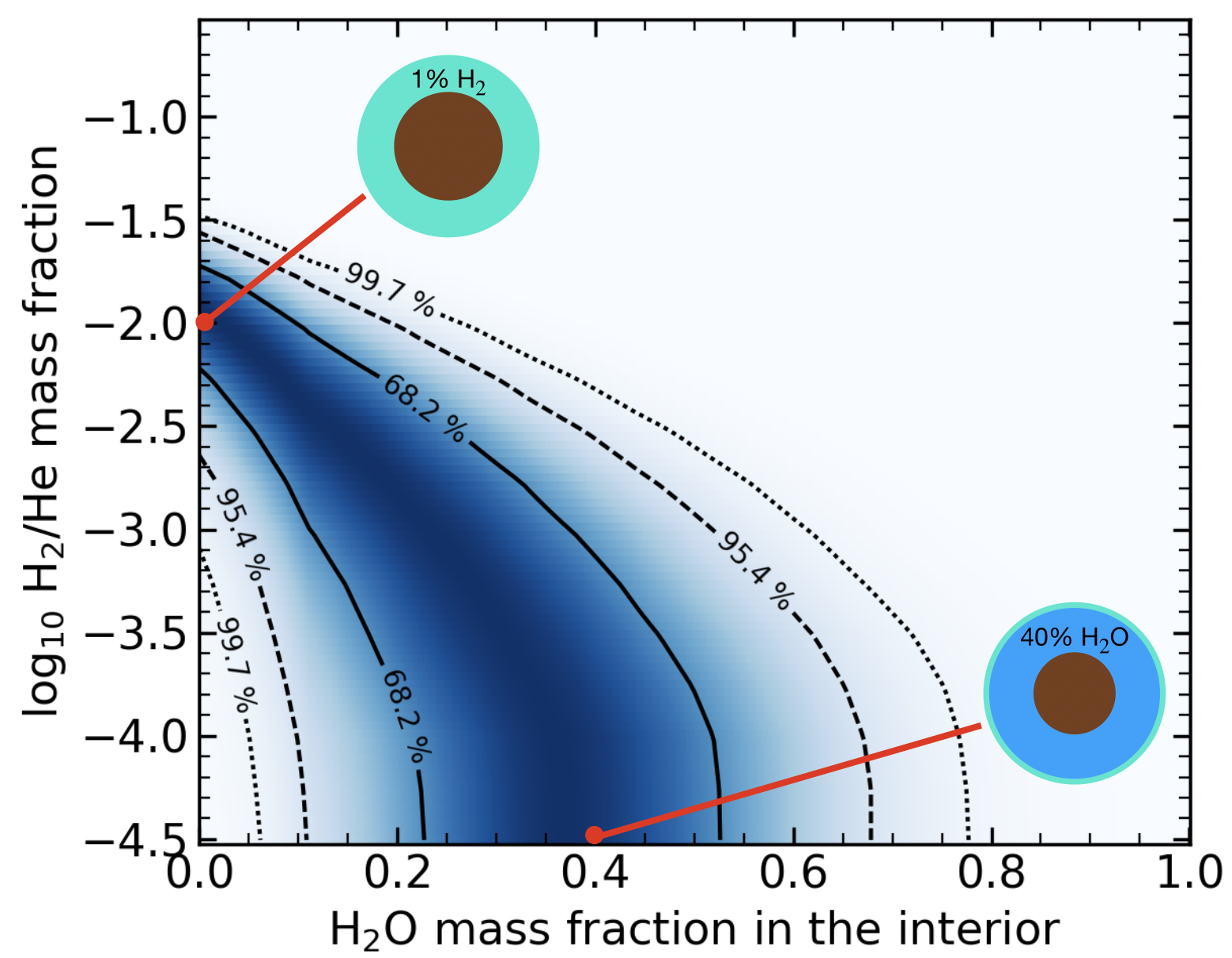}}
  \caption{\textbf{Left :} Mass-radius measurements of small exoplanets compared to theoretical curves from interior models for different compositions. TOI-824\,b is shown in yellow with 1-$\sigma$ error bars \citep{burt_toi-824_2020}; all other confirmed exoplanets with precise mass measurements ($<$ 20\%) are shown in light gray. Composition curves are shown for a range of interior compositions including 100\% iron, 100\% dry Earth-like interior and 100\% MgSiO$_3$ \citep[for T=1000\,K;][]{zeng_massradius_2016, zeng_growth_2019}; as well as for a 100\% water composition and for a rock/iron core with icy mantle (50\% Earth-like core + 50\% H$_2$O mantle) \citep[for T=1253\,K;][]{aguichine_massradius_2021}. In addition, scenarios with 1\% or 2\% H$_2$/He envelopes are shown for both the dry Earth-like interior as well as for the ``rocky core plus icy mantle" scenario using a temperature of 1000\,K \citep{zeng_massradius_2016, zeng_growth_2019}. The solar-system planet Neptune is added as a reference along with a hypothetical Neptune-like planet that lost its gaseous H$_2$/He envelope (core and mantle only). TOI-824\,b's mass and radius are consistent with an exposed Neptune mantle without H$_2$/He envelope (50\% Earth-like core + 50\% H$_2$O mantle) as well as with a dry Earth-like interior with 1\% H$_2$/He on the other extreme. \textbf{Right :} Two-dimensional joint posterior density distribution for the mass fraction of water in TOI-824\,b's core and mantle (horizontal axis) and the mass fraction of a possible H$_2$/He gaseous envelope on TOI-824\,b (vertical axis). The shading represents the probability density with the black curves indicating the 68.2\%, 95.4\%, and 99.7\% Bayesian credible regions. The interior water mass fraction ranges from 0 (core with Earth-like composition) to 1 (pure water interior). The mass fraction of the H$_\mathrm{2}$/He envelope is shown on a log scale, with negligible amounts of H$_2$/He towards the bottom of the diagram. Schematics illustrate the planet structure at the two extreme ends of the high probability region.}
  \label{fig:composition}
\end{figure*}
The recent \textit{TESS} discovery of the hot sub-Neptune TOI-824\,b provides us with a prime opportunity to go beyond transit spectroscopy and characterize the structure of a sub-Neptune via thermal emission measurements in secondary eclipse. TOI-824\,b is a sub-Neptune on a circular orbit around a K4V star with an effective temperature of 4600\,K \citep{burt_toi-824_2020}. With its size of 2.93\,$R_\mathrm{\oplus}$, this planet falls right in the sub-Neptune range. However, its mass of 18.47\,$M_\mathrm{\oplus}$ (measured through PFS and HARPS radial velocity observations) makes it a very dense planet \citep[4.03\,g/cm$^3$;][]{burt_toi-824_2020}, almost twice as dense as Neptune, and quite different from the lighter sub-Neptunes characterized in the past. Furthermore, the short 1.39-day orbit and hot zero-albedo full-heat-redistribution equilibrium temperature of TOI-824\,b (1253\,K) make it an ideal target for emission spectroscopy during eclipse: a method which is less affected by the presence of clouds in the atmosphere \citep{fortney_effect_2005, burt_toi-824_2020} and by star light contamination \citep{rackham_transit_2018} than transit spectroscopy. With this work, TOI-824\,b is among the smallest exoplanets and the first dense sub-Neptune for which the thermal emission is measured. 

On its short 1.39-day orbit, TOI-824\,b receives an extreme amount of radiation, which is $\sim$410 times more intense than the solar irradiation of Earth. Hence, it is likely that TOI-824\,b has experienced or is still experiencing important photoevaporation and atmospheric escape. However TOI-824\,b is thought to have kept its atmosphere over its lifetime because of its large mass \citep{burt_toi-824_2020}.



In this work, we combine an interior analysis of TOI-824\,b along with an atmosphere analysis of new \textit{Spitzer} secondary eclipse observations in order to provide evidence for TOI-824\,b being an exposed Neptune mantle with a metal-rich atmosphere. In Section \ref{sec:comp}, we provide a reanalysis of the possible bulk compositions of the planet using theoretical composition models. In Section \ref{sec:obs}, we describe the analysis of the \textit{Spitzer} observations obtained for this study. Section \ref{sec:atmosModel} presents the atmospheric analysis of the secondary eclipse measurements of TOI-824\,b and we discuss the main results of this work along with possible formation pathways for TOI-824\,b in Section \ref{sec:disc}. We end by presenting our conclusions and future observation prospects in Section \ref{sec:conc}.


\section{Revisiting the interior composition of TOI-824\,b}\label{sec:comp}
Sub-Neptunes are diverse. Despite its similar radius, TOI-824\,b has a much larger density than previously studied sub-Neptunes like GJ\,1214\,b \citep{kreidberg_clouds_2014} or K2-18\,b \citep{cloutier_characterization_2017, benneke_spitzer_2017, benneke_water_2019}. As we discuss in this section, TOI-824\,b occupies a position in mass-radius space which does not necessarily indicate a H$_\mathrm{2}$-rich atmosphere. We discover that the properties of TOI-824\,b are consistent with an ice-rich ``exposed Neptune mantle" exoplanet without a hydrogen envelope at one extreme, as well as with a rocky planet in a hydrogen envelope at the other extreme (Figure \ref{fig:composition}). 

\subsection{Interior models with self-consistent non-gray atmospheres}
Following the approach of C. Piaulet et al. (2022, in preparation), we explore the range of possible compositions of TOI-824\,b by connecting planet interior modeling \citep{thorngren_connecting_2019} with self-consistent non-gray modeling of the atmosphere (see Section \ref{sec:atmosModel}) to create full-planet models that can be compared to the mass and radius of TOI-824\,b. We create a grid of full-planet models ranging from 0.5 to 20\,M$_\mathrm{\oplus}$ with different mass fractions of the three following components: a rock/iron core (Earth-like composition), water, and H$_\mathrm{2}$/He. For each model in our grid, a SCARLET radiative-convective chemical equilibrium model for a solar composition (see Section \ref{sec:atmosModel} for more details) is connected on top of the interior model, while simultaneously ensuring that the boundary temperature (interface between the interior and the atmosphere) and the bulk composition is consistent (C. Piaulet et al. 2022, in preparation). We then use this model grid to predict a planet radius for all combinations of the planet mass, water mass fraction, and H$_\mathrm{2}$/He mass fractions. 

In this framework, we use a blackbody emission spectrum defined by the host star's effective temperature to model the stellar spectrum. The planetary radii predicted by the framework are mostly sensitive to the total energy budget of the planet. Hence, using a more realistic stellar model including absorption lines would likely have a negligible effect on the predicted planetary radii. We test this by comparing the radius predicted by our framework for a model using a blackbody stellar spectrum with another model using a PHOENIX-ACES stellar spectrum \citep{husser_new_2013}. The difference in the predicted radius between the two models is of 0.45\%, far below the actual uncertainty on the planet radius (6.8\%) and other model uncertainties (such as the choice of metallicity, etc.).

\subsection{Constraining TOI-824\,b's interior composition}
Using the grid of full-planet models, we constrain the interior water mass fraction and the mass fraction of a possible hydrogen envelope in a Bayesian analysis. We assign a uniform prior on the water fraction between 0 and 1 and take the remainder to be a rock/iron core with an Earth-like iron-to-rock ratio. For the possible gaseous H$_2$/He envelope, we assign a log-uniform prior between $10^{-5}$ and 0.3 in fraction of the planet's total mass, with values near the lower limit resulting in a negligible contribution to the planet's radius (Figure \ref{fig:composition}). We then assign the mass measurement of TOI-824\,b as Gaussian prior and evaluate the likelihood for each combination of planetary mass, interior water mass fraction, and H$_2$/He envelope mass fraction by comparing the measured radius of TOI-824\,b to the model radius interpolated from the full-planet model grid (C. Piaulet et al. 2022, in preparation).

The resulting 2D maginalized posterior distribution (Figure \ref{fig:composition}) reveals that the observed mass and radius of TOI-824\,b are consistent with bulk compositions along a curve in the water mass fraction/hydrogen mass fraction space. A water-rich mantle (38$\pm 15$\,\% by mass) on top of a rocky core agrees with the data with no hydrogen envelope, while a dry rock/iron interior would require a gaseous hydrogen envelope of approximately 1\% in mass. In between, a continuum of possible models can theoretically explain the observed properties of TOI-824\,b. We note that the water-rich scenario results in a composition similar to that of Neptune's mantle: a high-metallicity, water-rich interior. However, this composition does not allow for a significant hydrogen envelope ($<$ 0.01\%, see Figure \ref{fig:composition}); we henceforth refer to it as the ``exposed Neptune mantle'' scenario.  Moving forward, we will try to understand which of these compositions, the exposed Neptune mantle or the hydrogen-dominated envelopes, represents TOI-824\,b best, as constraining the exact composition of the planet is beyond the scope of this study.


\section{Analysis of the \textit{Spitzer} eclipses}\label{sec:obs}

\subsection{Observations and Data Reduction}
We obtained secondary eclipse observations of TOI-824\,b with the \textit{Spitzer Space Telescope} IRAC instrument in both Channel 1 and Channel 2. These observations were part of the large \textit{TESS} planets follow-up program (GO 14084, PI Crossfield). The hot temperature of TOI-824\,b is what allowed us to obtain a high enough signal-to-noise ratio (SNR) since the signal in secondary eclipse observations scales directly with the thermal emission of the planet which in turn depends on the planet's temperature. TOI-824\,b was observed eight times from November to December 2019, four secondary eclipses were observed in each channel (Table \ref{tab:obs}). Each eclipse observation was preceded and followed by one hour of observation to ensure enough baseline data to correctly model the light curve. We used 2-second exposures and the total observing time for the eight eclipses was of $\sim$24 hours.

We process the \textit{Spitzer}/IRAC images following standard methods. We start from the Basic Calibrated Data (BCD) images and follow the procedure detailed in \citet{benneke_spitzer_2017} and \citet{benneke_water_2019} for estimating the background flux and the centroid position of the star. When determining the stellar centroid, the aperture was reduced to exclude a stellar neighbour and avoid erroneous stellar position. This did not affect the rest of the reduction which was conducted as in \citet{benneke_water_2019}.

\subsection{Data Analysis}

\begin{table}
\begin{tabularx}{\columnwidth}{ccc}
\hline
\hline
& \textbf{Observations} & \\
\hline
Instrument & Wavelength ($\mu$m) & UT Start Date \\
\hline
\textit{Spitzer}/IRAC Ch1 & 3.15 -- 3.94 & 2019 Nov 22\\
&& 2019 Nov 29\\
&& 2019 Dec 01\\
&& 2019 Dec 02\\
\\
\textit{Spitzer}/IRAC Ch2 & 3.96 -- 5.02 & 2019 Dec 10\\
&& 2019 Dec 16\\
&& 2019 Dec 20\\
&& 2019 Dec 31\\
\hline
\hline
& \textbf{Measurements} & \\
\hline
Observable & Unit & Value\\
\hline
&\textbf{Channel 1}&\\
Eclipse depth & ppm & 142$^{+57}_{-52}$\\
& T$_{b}$(K) & 1463$^{+183}_{-196}$\\
$\sigma_\textrm{Nov 22}$ & ppm & 914.3$^{+31.1}_{-28.9}$\\
$\sigma_\textrm{Nov 29}$ & ppm & 909.4$^{+27.8}_{-30.4}$\\
$\sigma_\textrm{Dec 01}$ & ppm & 910.8$^{+33.0}_{-29.4}$\\
$\sigma_\textrm{Dec 02}$ & ppm & 893.2$^{+35.0}_{-28.4}$\\
&\textbf{Channel 2}&\\
Eclipse depth & ppm & 245$^{+75}_{-77}$\\
& T$_{b}$(K) & 1484$^{+180}_{-202}$\\
$\sigma_\textrm{Dec 10}$ & ppm & 1212.9$^{+39.3}_{-38.5}$\\
$\sigma_\textrm{Dec 16}$ & ppm & 1124.9$^{+38.3}_{-34.9}$\\
$\sigma_\textrm{Dec 20}$ & ppm & 1228.9$^{+40.3}_{-39.4}$\\
$\sigma_\textrm{Dec 31}$ & ppm & 1287.5$^{+41.7}_{-36.1}$\\
\hline
\hline
& \textbf{Orbital} & \\
&\textbf{Parameters}&\\
\hline
Parameter & Unit & Value\\
\hline 
T$_\mathrm{sec}$ & BJD$_\mathrm{TDB}$ & $2458849.247 ^{+0.007}_{-0.004}$\\
\\
$\phi_\mathrm{sec}$ & - & 0.49997$^{+0.00527}_{-0.00365}$\\
\\
$e \, \cos(\omega)$ & - & -0.00005$^{+0.00828}_{-0.00574}$\\
\hline
\hline
& \textbf{Atmospheric} & \\
&\textbf{Parameters}&\\
\hline
Parameter & 68.2\% Bayesian        & 95.4\% Bayesian\\
          & credible interval      &  credible interval\\
\hline 
Bond albedo (A$_\mathrm{B}$) & [0.0, 0.26] & [0.0, 0.49]\\
\\
Heat redist. & [0.49, 2/3] & [0.33, 2/3]\\
factor ($f$) & &\\
\hline
\hline
\end{tabularx}
\caption{\label{tab:obs} Summary of the secondary eclipse observations of TOI-824\,b with measurements and derived physical parameters. The eclipse depths are given both in ppm and in brightness temperature (T$_{b}$(K)). The fitted photometric scatter is also quoted for each visit. See Figure \ref{fig:spectrum} for the full 2D distribution of the atmospheric parameters.}
\end{table}

We perform the analysis of our eight eclipse light curves using the ExoTEP framework \citep{benneke_spitzer_2017, benneke_water_2019, benneke_sub-neptune_2019}. ExoTEP uses a Markov Chain Monte Carlo (MCMC) algorithm to fit the secondary eclipse light curve, the systematics models and the photometric noise parameters. For each \textit{Spitzer}/IRAC channel, we first perform individual MCMC analyses to each secondary eclipse data set and then perform a joint analysis of all of the four eclipses together. In our MCMC, the only astrophysical parameters that we are fitting are the eclipse depths, the time of the secondary eclipse and the photometric scatter in each visit. We use the well-constrained orbital parameters from \citet{burt_toi-824_2020} for the other needed physical parameters.

We use the well-known Pixel Level Decorrelation (PLD) model to account for the systematics associated with the movement of the star on the detector coupled with the sensitivity variations within the detector pixels \citep{deming_spitzer_2015, benneke_spitzer_2017}. We also use a systematics model which includes a second order logarithmic ramp in time. The analysis is performed on the full \textit{Spitzer} timeseries for each eclipse, and we initially bin the data by 20-second intervals. We do not include a phase-curve component to our light-curve modeling, as the data is tightly centered on the eclipse, and our systematics model already includes a ramp at the start of the observations, which would result in a degeneracy with the phase-curve model. Furthermore, if we compute a rough estimate of the variation in the signal due to the phase curve for two hours of observations (under the extreme assumptions that the night-side temperature is zero and the phase-curve variation is monotonic and linear over the full two hours), we find an expected signal around 5 ppm, far below the precision on our eclipse depths.

The secondary eclipse light-curve model we use in our MCMC analysis is generated using the Batman python package \citep{kreidberg_batman_2015}. As described earlier, we perform a joint fit of the light curves to obtain our best estimate of the eclipse depth while still fitting individual systematics models for each visit \citep{benneke_water_2019}. To speed up the process, the joint analysis is initiated with the best-fit parameters obtained from the individual eclipse fits for the systematics model and an average of the previous fits for the eclipse depth. We use the emcee python package \citep{foreman-mackey_emcee_2013} to perform all MCMC analyses and obtain the joint posterior distributions of the parameters, using uniform priors on each of them. 

\begin{figure*}[t!]
  \centering
  \subfigure{\includegraphics[width=0.48\textwidth]{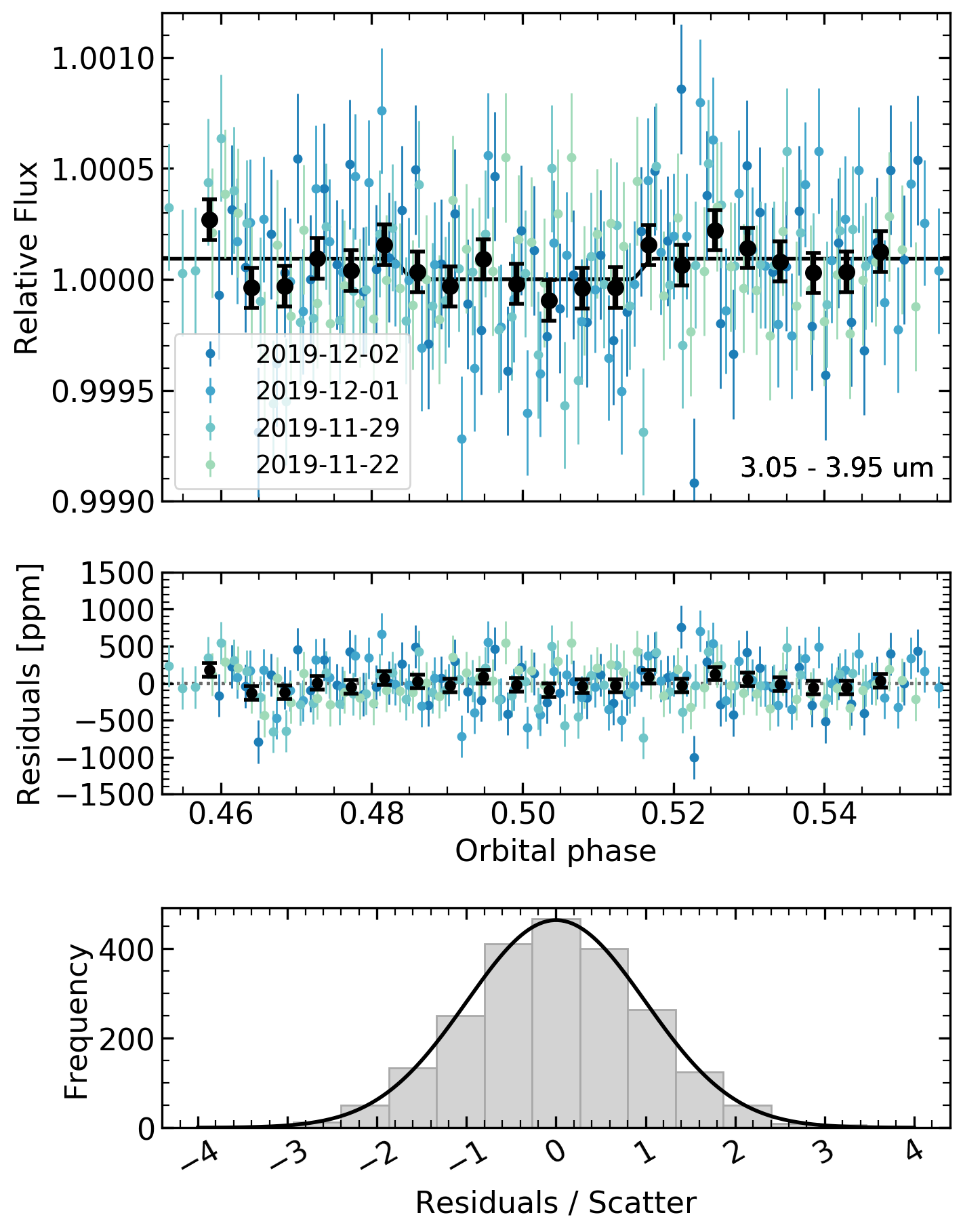}}
  \hfill
  \subfigure{\includegraphics[width=0.48\textwidth]{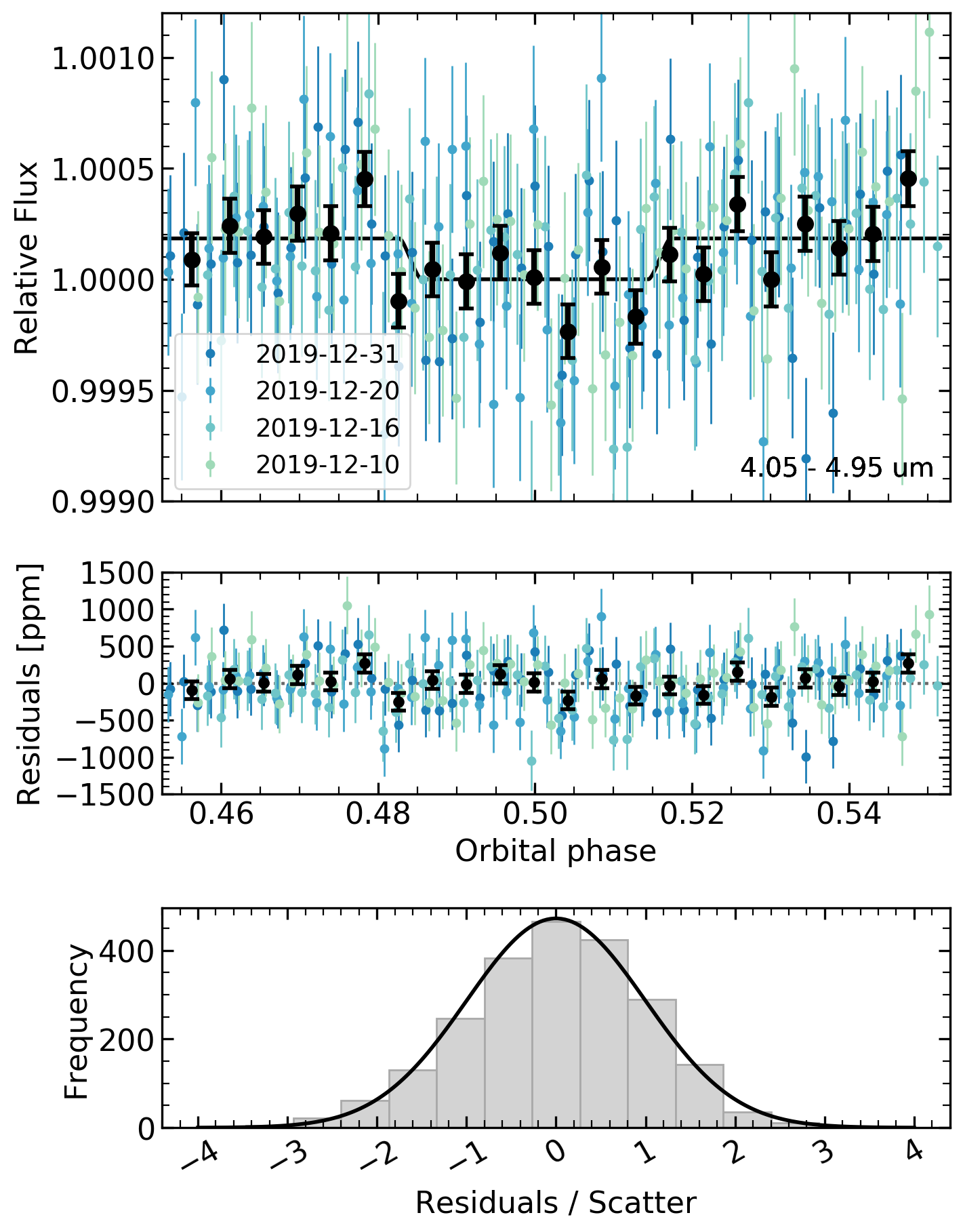}}
  \caption{White-light-curve fits from the joint analysis of each set of 4 \textit{Spitzer} eclipse observations of TOI-824\,b in Channel 1 \textbf{(left)} and Channel 2 \textbf{(right)}. The top panels show the best-fit light curves (black curve), with the systematics-corrected observations (blue circles) and their error bars. The middle panels show the residuals from the light-curve fits shown in the top panels. The \textit{Spitzer} data points are binned to 3.33-minute intervals and the bold black points show the combination of all the data binned by groups of 100 points. In the bottom panels, the residuals divided by the photometric scatter for each eclipse are shown as histograms along with the expected Gaussian curve. The error distribution follows the expected Gaussian distribution.}
  \label{fig:SpitzerWLCfit}
\end{figure*}

Our first ExoTEP analysis is performed only on Channel 2 (which has a stronger signal) and fits both the eclipse depth and the time of secondary eclipse in order to verify that the orbit is circular \citep{burt_toi-824_2020}. We use the time of secondary eclipse measured by our joint analysis for the time of the last visit of Channel 2 to provide an updated ephemeris for TOI-824\,b (Table \ref{tab:obs}). We use the posterior distribution on the time of secondary eclipse obtained from the joint analysis to derive constraints on the phase of the eclipse and the eccentricity term $e \, \cos(\omega)$ of TOI-824\,b's orbit. We take samples from our posterior on the time of secondary eclipse and convert them to orbital phase and to the $e \, \cos(\omega)$ term using samples for the period and transit time of TOI-824\,b, assuming Gaussian distributions defined by the values quoted in the discovery paper for both parameters \citep{burt_toi-824_2020}. This allows us to find that $e \, \cos(\omega) = -0.00005^{+0.00828}_{-0.00574}$ and to confirm that the orbit of TOI-824\,b is circular (Table \ref{tab:obs}).

We perform our ExoTEP analysis on both channels now assuming a circular orbit in order to obtain the eclipse depths. Hence, combining the astrophysical and systematics parameters, our joint analyses are fitting for the eclipse depth, the photometric scatter in each visit, the PLD coefficients in each visit and the second order logarithmic ramp terms in each visit. We investigate the scatter of the binned residuals of our resulting light curves and find that they follow the photon-noise expected trend of $1/\sqrt{N}$, showing no signs of correlated noise. The eclipse depth retrieved for Channel 2 is consistent with the result obtained from the previous fit where the time of eclipse was not fixed. Furthermore, we find that the precision on the eclipse depths we obtained is comparable to previous estimates of the expected precision on the eclipses. The joint white-light-curve fits along with the best-fit eclipse light curves for Channels 1 and 2 are shown in Figure \ref{fig:SpitzerWLCfit}. 

From our MCMC analysis of the light curve, we measure eclipse depths of 142$^{+57}_{-52}$ and 245$^{+75}_{-77}$\,ppm for TOI-824\,b in the 3.6 and 4.5\,$\mu$m \textit{Spitzer} channels. These eclipse depth measurements can be converted into brightness temperatures (considering a Phoenix stellar spectrum, see section \ref{sec:atmosModel}) and give 1463$^{+183}_{-196}$\,K and 1484$^{+180}_{-202}$\,K at 3.6 and 4.5\,$\mu$m respectively (Table \ref{tab:obs}). 


\section{Atmospheric Analysis}\label{sec:atmosModel}
In order to discover which of the compositions proposed in Section \ref{sec:comp} describes best TOI-824\,b, we produce multiple models of the atmosphere of TOI-824\,b that we subsequently compare to our observations. We use two different atmosphere modeling frameworks to model TOI-824\,b: the 1D self-consistent atmosphere modeling framework SCARLET and the 3D GCM simulation framework SPARC/MITgcm. In this section, we describe the methods and the results associated with each framework.

\begin{figure*}[t!]
  \centering
  \subfigure{\includegraphics[width=0.54\textwidth]{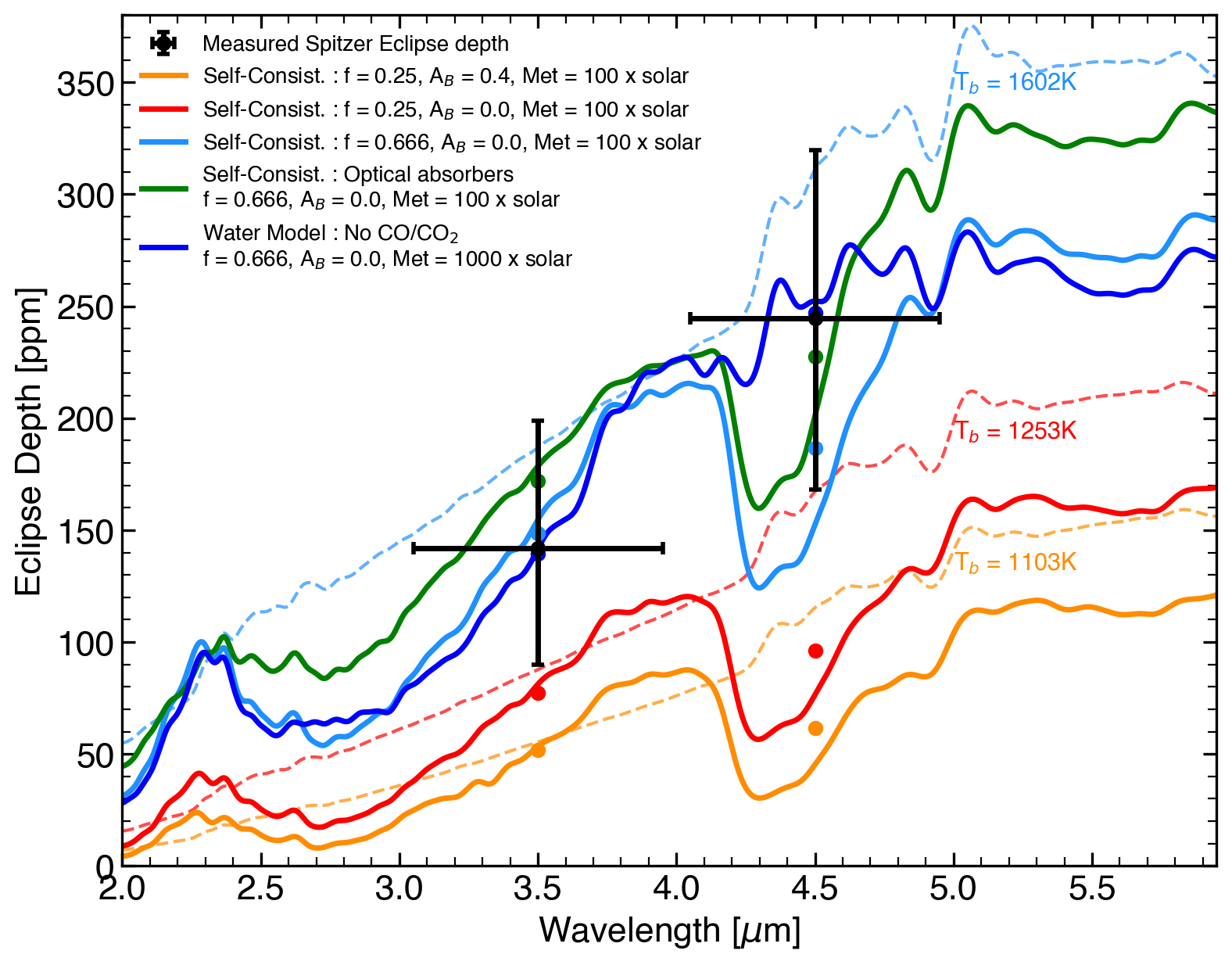}}
  \subfigure{\includegraphics[trim = 0 0 40 60, clip, width=0.44\textwidth]{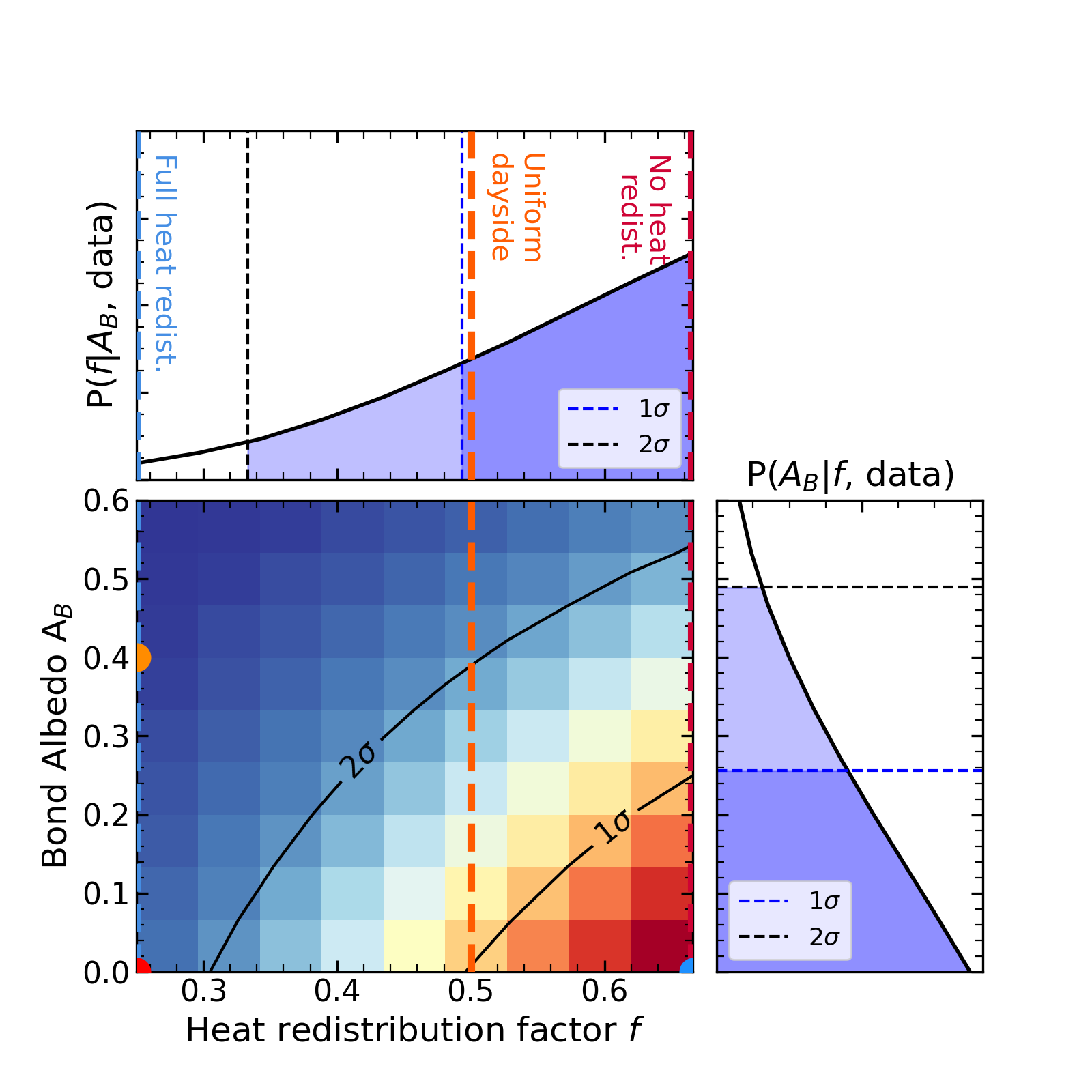}}
  \caption{\textbf{Left:} \textit{Spitzer} secondary eclipse depth measurements (black points) compared to model secondary eclipse spectra of TOI-824\,b. Dashed lines show blackbody eclipse spectra for different day-side brightness temperatures corresponding to different combinations of low or high Bond albedos ($A_\mathrm{B}=0$ and $A_\mathrm{B}=0.4$) and complete or poor heat redistribution across the surface of the planet ($f=0.25$ and $f=0.666$). Theoretical model eclipse spectra based on self-consistent 1D atmospheres for equivalent low or high albedos and full or poor heat redistributions are shown as solid lines in matching colors (orange, red, light blue). In addition, a model including short-wavelength absorbers TiO, SiO and VO (green) and a water-dominated case where CO and CO$_2$ are removed (blue) are shown. The absorption features around 4.5 $\mu$m are driven by CO and CO$_2$ opacities. Colored circles represent the models integrated over the \textit{Spitzer} bandpasses. \textbf{Right:} Joint posterior distribution of the Bond albedo (vertical axis) and heat redistribution factor (horizontal axis) for the grid of 1D self-consistent atmosphere models. The 2D posterior is shown as colored shading, with the 1$\sigma$ and 2$\sigma$ credible region indicated by black lines. Colored circles in the posterior correspond to models shown in matching colors in the spectrum. The top and right panels show the marginal distributions with the 1$\sigma$ and 2$\sigma$ limits of the heat redistribution factor and Bond albedo, respectively. Colored dashed lines are added along the heat redistribution axis to highlight special values for the parameter ($f$=0.25, 0.5, and 2/3). } 
  \label{fig:spectrum}
\end{figure*}

\subsection{Grid of SCARLET 1D self-consistent models}
\subsubsection{Methods}
We use the SCARLET atmosphere modeling framework \citep{benneke_atmospheric_2012, benneke_how_2013, benneke_strict_2015, benneke_water_2019, benneke_sub-neptune_2019, pelletier_where_2021} in order to create a grid of 1D self-consistent atmosphere models of TOI-824\,b and compare the model eclipse spectra to the \textit{Spitzer} eclipse measurements. We use the framework to produce chemical equilibrium, radiative, non-gray models of TOI-824\,b. These are wavelength-dependent models that solve the equations of chemical equilibrium and radiative transfer iteratively, and produce a theoretical temperature-pressure profile of the planet. Once the model converges to a stable temperature structure and chemistry, it computes the associated secondary eclipse spectrum. We account for the stellar spectrum in both the secondary eclipse spectrum calculation and the temperature structure calculation based on a Phoenix stellar model \citep{husser_new_2013, lim2015pysynphot} (T$_\mathrm{eff}$ = 4600\,K, log$\,g$ = 4.605, [M/H] = 0). 

This work focuses on constraining two dimensionless parameters which affect the energy budget of TOI-824\,b: the Bond albedo $A_\mathrm{B}$ and the heat redistribution factor $f$. The former is the total (i.e., integrated over all wavelengths) reflectivity index of the planet and the latter represents the efficiency of the heat transport around the planet \citep{hansen_absorption_2008, seager2010exoplanet}. The Bond albedo can range from zero (no reflected light) to one (all light is reflected), whereas the heat redistribution factor can range from 0.25 (heat uniformly distributed around the planet) to $\frac{2}{3}$ (no heat transport, the planet emits light like a bare rock, with more emission near the subsolar point) \citep{hansen_absorption_2008, seager2010exoplanet}. Another special case for this heat redistribution factor $f$ is 0.5, which represents uniformly distributed heat, but only on the day side of the planet \citep{hansen_absorption_2008}. The SCARLET framework takes both of these parameters as inputs in order to compute the equilibrium temperature $T_\mathrm{eq}$ of the 1D models via 
\begin{equation}\label{eq:Teq}
    T_\mathrm{eq} = T_\mathrm{eff,*} \left(\frac{R_*}{a}\right)^{1/2} \left[f(1 - A_\mathrm{B})\right]^{1/4},
\end{equation} where $T_\mathrm{eff,*}$ is the stellar effective temperature, $R_*$ is the stellar radius and $a$ is the semi-major axis of the orbit \citep{seager2010exoplanet}. Hence, we create a grid of TOI-824\,b models for different values of these parameters and find which models are consistent with the observed eclipse depths. 

\subsubsection{Results}
A Bayesian analysis of the data with our 1D atmosphere grid shows that the \textit{Spitzer} observations indicate a poor heat redistribution around the planet and a small Bond albedo. We use Bayes' Theorem to compute the posterior probability distribution of our grid of models using uniform priors and the \textit{Spitzer} measurements assuming Gaussian noise. We produce the models assuming chemical equilibrium, solar carbon-to-oxygen ratio and 100 $\times$ solar metallicity. For time purposes, we use a grid of 100 models, exploring 10 values for each of the parameters. For the Bond albedo, we restrict the values from 0.0 to 0.6, as values approaching one (full reflection) would be unphysical. We let the heat redistribution factor explore the whole parameter space from 0.25 to 0.666. The resulting two-dimensional posterior distribution favors a poor heat transport to the night side and a low Bond albedo (Figure \ref{fig:spectrum}), again favoring a hot day-side temperature. For instance, assuming a planetary albedo of zero, comparing the hot no-heat-redistribution model with $f$=0.666 to the full-heat-redistribution model with $f$=0.25 yields a Bayes factor of 10.5 in favor of the model without heat transport.

Although the two \textit{Spitzer} points do not have a high enough resolution to allow us to constrain molecular features in the data, the Channel 2 measurement is deeper than the 4.5\,$\mu$m absorption in our models (Figure \ref{fig:spectrum}), which is driven by CO and CO$_2$ opacities. In order to get better consistency with the Channel 2 measurement, we need to include optical light absorbers like TiO and SiO in the atmosphere (which is done self-consistently and still under the chemical equilibrium assumption) to try and induce absorption in the upper atmosphere (Figure \ref{fig:spectrum}, green model). We also produce a water-dominated toy model by increasing the metallicity, decreasing the carbon-to-oxygen ratio, and removing CO and CO$_2$ from the model. Although this breaks the chemical equilibrium, we find that the 4.5\,$\mu m$ absorption disappears and that the model is consistent with the measurements. Higher-resolution observations will be needed to constrain molecular abundances, but our measurements favor models without strong 4.5\,$\mu$m absorption (water-dominated or with optical absorbers) to models with CO/CO$_2$-driven absorption.

It is not surprising that both the heat redistribution factor and the Bond albedo are not tightly constrained as these parameters are strongly correlated. Indeed, increasing the heat redistribution factor $f$ or decreasing the Bond albedo $A_\mathrm{B}$ both have the same effect of raising the equilibrium temperature, as can be seen in equation \ref{eq:Teq}. This effect also appears in the posterior distribution (Figure \ref{fig:spectrum}) which displays a positive correlation between the parameters. Furthermore, we see that the marginal probability distributions for the Bond albedo and heat redistribution factor are hugging the boundaries of the parameter space allowed by our priors (Figure \ref{fig:spectrum}). However, we recall that we used uniform priors on the range of physical and possible values on both parameters. Hence, the shapes of the marginal distributions are merely a consequence of the fact that the high-probability region is close to the extreme physical values for the parameters and that the \textit{Spitzer} eclipse depths are not precise enough to rule out the borders of the parameter space. 

We also investigate the constraints on the optical eclipse that can be obtained via the TESS photometry, which could potentially help constrain the Bond albedo with better accuracy. However, the expected eclipse signal in the TESS band is much smaller ($\sim$13\,ppm) than the precision obtained by the TESS observations ($\sim$80\,ppm), making it impossible to gain optical eclipse constraints. Hence, since the combination of a high heat redistribution factor and low Bond albedo is favored by our Bayesian analysis, we conclude that the data suggests a hot day-side temperature and a poor heat transport on TOI-824\,b. 

\subsection{3D General Circulation Models}
\subsubsection{Methods}
In addition to the grid of 1D models, we also use the 3D atmosphere modeling framework described in \citet{parmentier_thermal_2018} to create 3D simulations of the evolution and dynamics of the atmosphere of TOI-824\,b. This framework combines the SPARC/MITgcm general circulation model \citep{showman_atmospheric_2009}, the radiative transfer code from \citet{marley1999thermal}, and a modified NASA CEA Gibbs minimization framework to solve for the chemistry in the atmosphere \citep{gordon1994computer}. The GCM module solves the primitive equations of fluid dynamics on a cubic-sphere geometry, while the radiative transfer is solved using the two-stream approximation \citep{toon_rapid_1989}. Hence, the framework solves simultaneously for the atmosphere circulation, radiation, and chemistry. We produce the simulations for different values for the metallicity of the atmosphere in order to study how the secondary eclipse spectrum varies with the metallicity. Once the GCM has integrated the model over multiple days (earth days) and converges to a state where the overall trends in the photosphere do not change anymore, we take a snapshot of the planet on the last day of the simulation and from the day-side emission spectrum, we obtain the associated model secondary eclipse spectrum, which we can compare to our \textit{Spitzer} measurements. 

\begin{figure}[t]
\includegraphics[width=\linewidth]{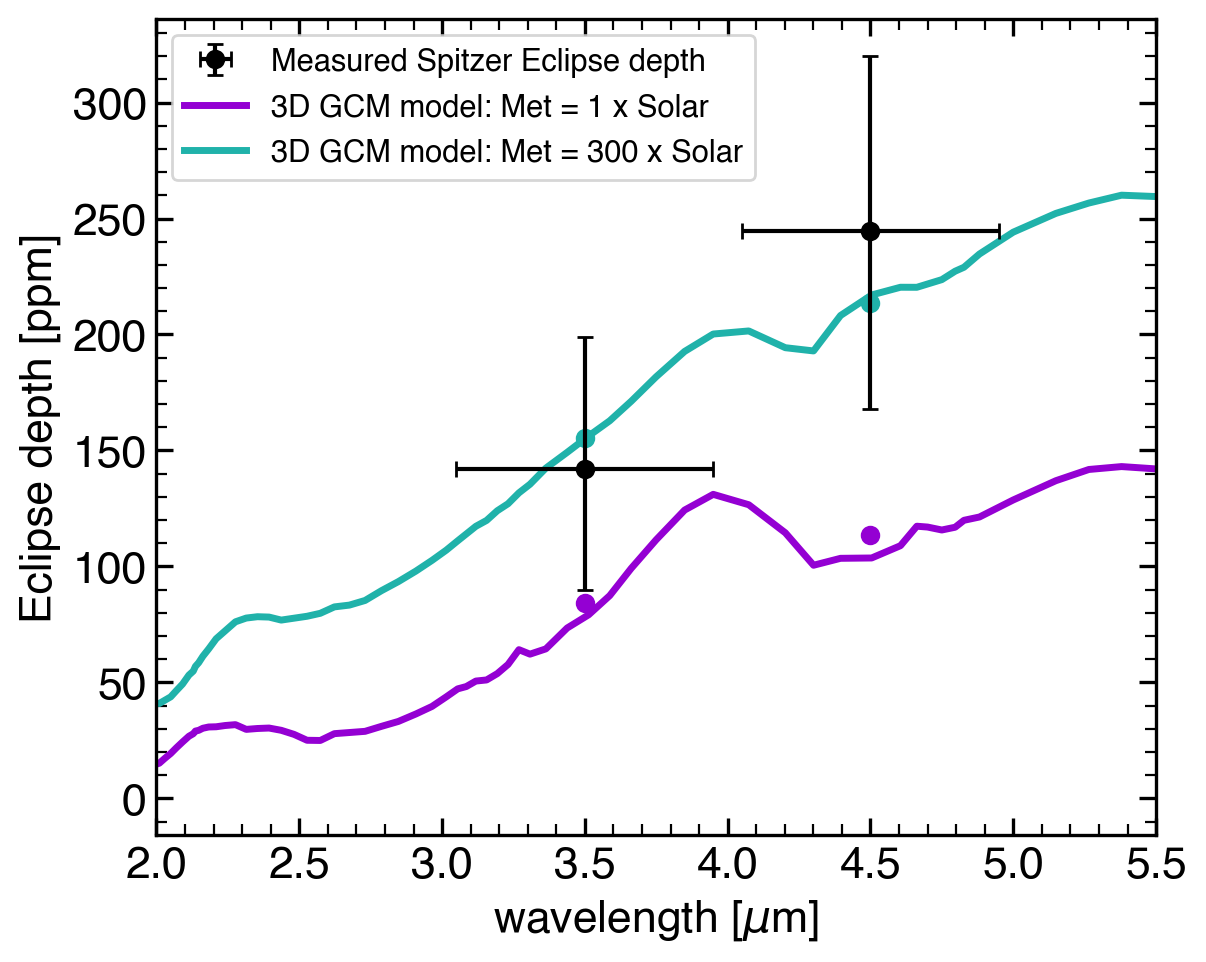}
\caption{\textit{Spitzer} secondary eclipse depth measurements (black points) compared to model secondary eclipse spectra of TOI-824\,b from the 3D GCM. The purple line shows an atmosphere with a solar composition while the blue line shows an atmosphere for 300 times the solar composition. The colored points show the GCM models integrated over the \textit{Spitzer} bandpasses. Both models were integrated over a period of 250 days and the spectrum shows a snapshot of the last day of the simulation. The \textit{Spitzer} measurements are in agreement with the metal-rich model and rule out the metal-poor model, suggesting that TOI-824\,b is consistent with a metal-rich atmosphere.}
\label{fig:GCM}
\end{figure}

\subsection{Results}
We produce two 3D GCM simulations of TOI-824\,b in order to model the scenarios described in section \ref{sec:comp}: one model with solar metallicity (hydrogen-dominated envelope) and one model with 300 times the solar metallicity (exposed Neptune mantle). We choose the value of 300 times the solar composition as it is the highest metallicity allowed by the framework. The initial temperature profile is calculated from the \citet{parmentier_non-grey_2015} analytical model using an equilibrium temperature of 1332\,K, which is slightly hotter than the equilibrium temperature of TOI-824\,b, to ensure a faster convergence of the GCM thermal structure. The time step in the numerical integration of both models is of 10 seconds and they both run for a total of 250 days. The final heat capacities and mean molecular weights of the atmospheres are consistent with the respective metallicities. From both models, we take a snapshot of the planet on the last day of the simulation and produce a secondary eclipse spectrum. 

The comparison of the eclipse depths of TOI-824\,b with GCM simulations suggests a high-metallicity atmosphere on TOI-824 b. Figure \ref{fig:GCM} shows the secondary eclipse spectra produced by both GCM simulations along with the \textit{Spitzer} eclipse measurements. The high-metallicity model displays a much larger eclipse signal which is consistent with both measurements, while the low-metallicity model, is not consistent with the measurements. Hence, the \textit{Spitzer} observations favor the high-metallicity model with a likelihood ratio of 7:1.

\section{Discussion}\label{sec:disc}
TOI-824\,b is a rare example of a sub-Neptune favorable to secondary eclipse observations that can give us key insights into the nature and diversity of the overall sub-Neptune population, especially in the regime of high-density sub-Neptunes. In this work, we investigated possible compositional scenarios for TOI-824\,b and we measured its day-side thermal emission with \textit{Spitzer} eclipse observations. Our results suggest that TOI-824\,b may not host a hydrogen-rich atmosphere and may therefore be substantially different in nature from our standard image of sub-Neptunes, often regarded as ``gas-rich'' super-Earths with voluminous hydrogen-envelopes \citep{schlichting_formation_2014, owen_evaporation_2017,bean_nature_2021}. The conclusion would be that within the overall sub-Neptune population, different categories of sub-Neptunes exist, or at least, that exceptions to the standard sub-Neptune image of a hydrogen-rich super-Earth exist.

\subsection{Observational hints at the exposed-Neptune-mantle scenario}
Our argumentation starts with a reanalysis of the interior composition of TOI-824\,b, which reveals that the range of compositions consistent with the measured mass and radius of TOI-824\,b extends from a super-Earth core with a hydrogen envelope, as suggested in the discovery paper \citep{burt_toi-824_2020}, all the way to a water-rich world without any hydrogen, reminiscent of an exposed Neptune mantle composition (Figure \ref{fig:composition}). While completely degenerate in mass and radius, the presence or lack of a hydrogen-dominated atmosphere on TOI-824\,b result in vastly different predictions for the planet's day-side thermal emission. A water-dominated envelope, on one hand, shows a characteristic poor atmospheric heat redistribution because the large abundances of strongly absorbing molecules increase the opacity of the atmosphere and raise the photosphere to lower pressures where the radiative timescale is shorter. As a result, the heat received from the host star is quickly re-radiated from the day side and is not transported around the planet. This results in a strong temperature contrast between the day and night sides of the planet and, consequently, a hotter day-side temperature \citep{menou_atmospheric_2012, kataria_atmospheric_2014, charnay_3d_2015, drummond_effect_2018}. A low-metallicity hydrogen-dominated envelope, on the other hand, has substantially lower mixing ratios of strong absorbers and thus a lower photosphere, which results in a stronger heat redistribution and therefore, a cooler day side (Section \ref{sec:atmosModel}).


To distinguish these scenarios, we present eight \textit{Spitzer} secondary eclipse observations that detect and characterize the thermal emission of TOI-824\,b's day side. Comparing the observed eclipse depths to the predictions from GCM simulations, we find that it is the metal-rich atmosphere scenario that agrees best with the \textit{Spitzer} measurements. Equivalently, using 1D atmosphere models, we find that the heat redistribution around the planet must be low ($f>0.49$), even when we consider the whole range of plausible Bond albedos between 0 and 0.6. Therefore, both the comparisons to the GCM and the 1D models hint at the high-metallicity atmosphere, which is consistent with the exposed Neptune scenario. In this context, it should be pointed out that this conclusion does not depend on the detailed model assumptions made in the GCM or the 1D models. The strong dependency of the heat redistribution on the amount of metals in the atmosphere has been shown in many independent studies of sub-Neptunes using different models \citep{menou_atmospheric_2012, kataria_atmospheric_2014, charnay_3d_2015, drummond_effect_2018}.

\subsection{Possible formation scenarios for TOI-824\,b}
\subsubsection{Close-in exoplanets formation pathways}
The current understanding of the formation of close-in exoplanets identifies two main scenarios that could lead to the formation of sub-Neptunes, which are often referred to as the ``drift model'' and the ``migration model'' \citep{bean_nature_2021}. While the drift model predicts the formation of rocky cores via pebble accretion inside of the ice line and close to the host star \citep{johansen_forming_2017}, the migration model predicts the formation of planetary cores outside of the ice line \citep{lambrechts_forming_2014} and their subsequent migration towards their short-orbit close-in position. In both models, the later stages of the formation include planetary cores colliding \citep{bean_nature_2021} and both models can lead to the formation of planetary bodies of similar size and mass.

While degenerate in mass-radius space, the planets formed via the drift or migration models will show characteristic differences in their bulk composition. In the drift model, the pebbles that create the planet core are thought to lose their metal and water contents as they drift in the disk and thus form rocky planets with an expected low amount of heavy volatiles like water \citep{bitsch_rocky_2019, bean_nature_2021}. In the migration model, large icy and rocky cores migrating from beyond the ice line are thought to keep most of their heavier volatiles during their migration, and thus form planets with metal-rich and water-rich envelopes \citep{bitsch_dry_2021, bean_nature_2021}. Now, if TOI-824\,b indeed lacks a low-metallicity, hydrogen-dominated gas envelope, as hinted at by the \textit{Spitzer} data, then we conclude that TOI-824\,b has a high volatile mass fraction of 38$\pm$15\,\% (Section \ref{sec:comp}), which could indicate that it formed via the migration model.

Other planet formation studies have explored the possibility of the close-in formation of water-rich planets (up to $\sim$20\% by mass) via degassing of the planet \citep{elkinstanton_ranges_2008}, or by invoking the need for an outer gas giant sparking an inward diffusion of water vapor into the disk inside of the ice line \citep{bitsch_dry_2021}. However, degassing  would not be able to explain a $\sim$40\% water content, and no evidence of a companion gas giant was found in the TOI-824 system so far. Moreover, following \citet{murphy_another_2021}, we compute a minimum mass enhancement factor for the protoplanetary disk compared to the minimum mass solar nebula to estimate how massive the protoplanetary disk would need to be in order to explain the in situ formation of TOI-824\,b (on its current orbit) \citep{schlichting_formation_2014}. We find that, in order for TOI-824\,b to form in situ, this mass enhancement factor must be of $\sim$250, even larger than for other studies of dense sub-Neptunes \citep[e.g.,][]{murphy_another_2021} and making it an extraordinarily massive disk when compared to studies of close-in super-Earth systems \citep{chiang_minimum-mass_2013}. Hence, we conclude that TOI-824\,b most likely formed via the migration model, which can explain the composition, the close-in orbit and the inferred high-metallicity envelope of the planet in a more natural way.

\subsubsection{Formation of an exposed Neptune mantle}
If TOI-824\,b indeed lacks a hydrogen-dominated, low-metallicity atmosphere, then it must imply that either TOI-824\,b never accreted a substantial hydrogen envelope or that it did accrete a hydrogen envelope but lost it over the evolution of the planet. Both scenarios require further discussion.

If TOI-824\,b never accreted a hydrogen envelope, this would mean that the planet's core and mantle alone have a mass of 18.5\,M$_\oplus$, substantially above the ``critical'' core mass of 10\,M$_\oplus$ at which the planetary cores are expected to trigger runaway gas accretion and create gas giants in gas-rich environments \citep{lee_boundary_2019}. One would then have to postulate that the massive core of TOI-824\,b never accreted a hydrogen envelope because the formation of the planet happened late in the life of the protoplanetary disk, in a gas-poor environment, just before the disk disappeared \citep{lee_breeding_2016, lee_boundary_2019}. Only then could one explain why such a massive core did not enter the runaway accretion phase and did not accrete a hydrogen envelope. 

If, alternatively, we assume that TOI-824\,b accreted a large hydrogen envelope earlier in its evolution, then we are confronted again to its large mass, which makes the planet almost immune to photoevaporative mass-loss. As it was shown in the discovery paper \citep{burt_toi-824_2020}, TOI-824\,b's gravitational binding energy is greater than the total XUV irradiation it has received from its star over its lifetime and the planet is thus likely to have survived atmospheric escape. We further investigate this by running VPLanet \citep{barnes_vplanet_2020} photoevaporation simulations for fiducial TOI-824\,b hydrogen-rich scenarios in which we add hydrogen layers of 1\% and 10\% by mass onto the 18.5\,M$_\oplus$ core and assume the current orbit of TOI-824\,b over eight billion years. In both cases, using standard photoevaporation assumptions, we would conclude that the atmospheric mass-loss is minimal and most of the atmosphere survives. A recent study of the ultra dense 40\,M$_\oplus$ Neptune-size planet TOI-849\,b, which investigated the coupled evolution of the orbital dynamics of the system and photoevaporation of the planet for different stellar rotation rates for the host star, showed that this 40\,M$_\oplus$ core could rapidly lose its large hydrogen envelope due to photoevaporation \citep{pezzotti_key_2021}. However, TOI-849\,b is closer than TOI-824\,b to its brighter G-type star, and the evolution models assumed an in situ or early migration formation, which is less consistent with a volatile-rich planet like TOI-824\,b. Hence, it is unlikely that photoevaporation alone would explain how a hydrogen layer on TOI-824\,b was lost. 

Atmospheric mass-loss via giant impacts could be an alternative explanation. Recent studies have shown that sub-Netptunes can be completely stripped of their primordial H/He atmospheres by giant impact events \citep{biersteker_atmospheric_2019}. The giant impact of planetary cores releases a lot of thermal energy that can drive an important atmosphere mass-loss, stripping partially or even totally the hydrogen envelope of the planet. Furthermore, giant impact between planetary cores are expected to happen in both the drift and migration models discussed earlier. This scenario would thus allow the massive core and interior of TOI-824\,b to accrete a substantial hydrogen envelope beyond the ice line, while also explaining how it can be found with a hydrogen-poor exposed Neptune composition today.

\begin{figure}[t]
\includegraphics[width=\linewidth]{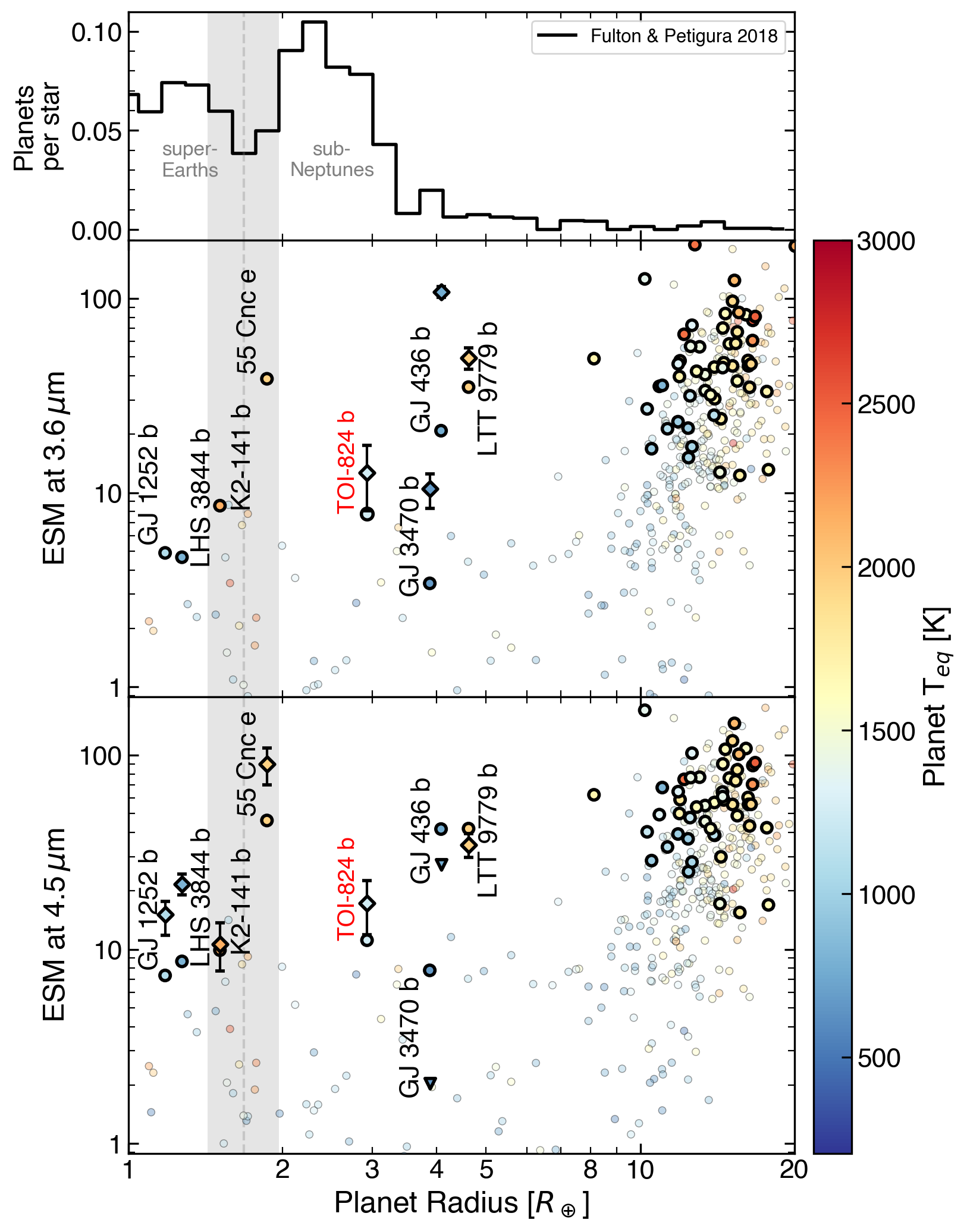}
\caption{\textbf{Top:} Occurrence rates of planets with orbital period smaller than 100 days, adapted from \citet{fulton_california-_2018}. \textbf{Bottom panels:} Emission Spectroscopy Metrics (ESM) \citep{kempton_framework_2018} at 3.6 and 4.5\,$\mu$m computed from the equilibrium temperature for confirmed exoplanets (circles). Exoplanets with reported secondary eclipse measurements are shown in bold \citep{deeg_characterization_2018}. Small planets with available \textit{Spitzer} 3.6 and 4.5\,$\mu$m thermal emission measurements are labeled and are shown with the reported \textit{Spitzer} day-side brightness temperatures (diamonds with error bars)\citep[][I. Crossfield et al. 2022, in prep.]{demory_variability_2016, kreidberg_absence_2019, crossfield_phase_2020, zieba_k2_2022}. For GJ\,436\,b and GJ\,3470\,b at 4.5 $\mu$m, we show the brightness temperature upper bounds as triangles \citep{stevenson_possible_2010, benneke_sub-neptune_2019}. The gray area in all panels highlights the radius valley. TOI-824\,b has the highest ESM in the sub-Neptune range between 2 and 4 Earth radii in both bandpasses.}
\label{fig:eclipse_pop}
\end{figure}

\section{Conclusions}\label{sec:conc}
In conclusion, we have combined the results from our atmosphere analysis of the \textit{Spitzer} observations of TOI-824\,b with a renewed interior analysis of the planet to provide a first line of evidence for TOI-824\,b being an exposed Neptune-like mantle. In fact, the possible compositions of TOI-824\,b highlight the fact that the sub-Netpune population is maybe not uniform, and can be divided into sub-categories of planets that have or that lack a large hydrogen envelope. This idea is reinforced by other recent works that identified and studied dense sub-Netpunes, like TOI-849\,b \citep{armstrong_remnant_2020, pezzotti_key_2021} or K2-182\,b \citep{murphy_another_2021}. This small population of recently discovered super-dense sub-Neptunes \citep{murphy_another_2021} suggests that TOI-824\,b might not be unique, and might be part of a sub-class of ``exposed mantle" sub-Neptunes. However, among these dense sub-Neptunes, TOI-824\,b stands out as the only object for which we have thermal emission measurements, which, as it turns out, do support an exposed-Neptune-mantle composition.

Further characterization of sub-Neptunes (and super-dense sub-Neptunes) are needed to untangle their composition and understand their formation, as the evolution pathways leading to either the hydrogen-rich or hydrogen-poor compositions are numerous. For instance, obtaining constraints on molecular abundances in sub-Neptune envelopes could help us constrain the carbon-to-oxygen ratios of these planets, a tell-tale sign of their formation location. Our detection of TOI-824\,b's hot day side identifies the planet as an exquisite target to study these questions via further emission spectroscopy (Figure \ref{fig:eclipse_pop}).

With the arrival of JWST, further spectroscopic eclipse observations of TOI-824\,b can be used in the coming years to test our hypothesis and confirm or disprove the ``exposed Neptune mantle'' nature of TOI-824\,b. In Figure \ref{fig:eclipse_pop}, we compare the Emission Spectroscopy Metric (ESM) \citep{kempton_framework_2018} of known exoplanets at 3.6 and 4.5\,$\mu$m. The ESM is a metric which yields high values for high SNR targets that are most favorable for emission spectroscopy. For all planets, we show the ESM assuming the zero-albedo and full-heat-redistribution equilibrium temperature (circles). For small exoplanets with \textit{Spitzer} eclipse observations, we also show the ESM based on the measured day-side brightness temperatures in each channel. Among the planets near the planet-radius peak of the sub-Neptune population, TOI-824\,b stands out as the sub-Neptune with the highest ESM in both the 3.6 and 4.5~$\mu$m bandpasses. TOI-824\,b thus represents a key target to follow-up in order to compare its composition to those of low-density (sub-)Neptunes (e.g., GJ\,3470\,b, GJ\,436\,b, K2-18\,b, LTT\,9779\,b), ultra-dense sub-Neptunes (e.g., TOI-849\,b, K2-182\,b) and smaller super-Earths (e.g., LHS\,3844\,b). Such follow-up observations might confirm the ``exposed mantle" nature of TOI-824\,b and establish the exposed Neptune mantles as a separate type of sub-Neptunes.

\acknowledgements

We wish to thank the reviewer for the provided comments, which greatly enhanced the manuscript. We further thank L.-P. Coulombe, S. Delisle and S. Pelletier for their insightful comments and ideas on the multiple iterations of the paper.

 P.-A. R. acknowledges financial support by the Natural Sciences and Engineering Research Council (NSERC) of Canada, the Fond de Recherche Qu\'{e}b\'{e}cois—Nature et Technologie (FRQNT; Qu\'{e}bec), the University of Montreal and the NSERC CREATE Technologies for Exo-Planetary Science (TEPS) program. D. D. acknowledges support from the TESS Guest Investigator Program grant 80NSSC19K1727 and NASA Exoplanet Research Program grant 18-2XRP18\_2-0136. 

This work is based on observations made with the \textit{Spitzer Space Telescope}, which is operated by the Jet Propulsion Laboratory, California Institute of Technology under a contract with NASA. 

\typeout{}
\bibliography{ZoteroBiblio,extra}



\end{document}